\documentclass[a4paper,11pt]{article}
\pdfoutput=1
\usepackage{jcappub}
\usepackage{amsmath,amssymb,graphicx,xcolor,hyperref}
\usepackage{mathtools,bm,booktabs,array,tabularx}
\usepackage[T1]{fontenc}
\usepackage{orcidlink}
\hypersetup{colorlinks=true,citecolor=blue,linkcolor=blue,urlcolor=blue,
 pdftitle={Exothermic dark-photon dark matter and the LZ high-recoil event: finite-reheating constraints},pdfauthor={Kuldeep Deka}}
\newcommand{\dd}{\mathrm{d}}
\newcommand{\Trh}{T_{\rm rh}}
\newcommand{\Tmax}{T_{\max}}
\newcommand{\Aprime}{A'}
\newcommand{\fH}{f_H}
\newcommand{\mpl}{M_{\rm Pl}}
\newcolumntype{Y}{>{\raggedright\arraybackslash}X}
\title{Exothermic dark-photon dark matter and the LZ high-recoil event: finite-reheating constraints}

\author{Kuldeep Deka \orcidlink{0000-0002-0064-5488}}
\affiliation{New York University Abu Dhabi\\
PO Box 129188, Saadiyat Island, Abu Dhabi, United Arab Emirates\\
and\\
Ruđer Bošković Institute, 10000 Zagreb, Croatia}
\emailAdd{kuldeepdeka.physics@gmail.com}

\abstract{An excited dark-matter state can release energy in a nuclear
collision and produce recoils well above the usual search window.
Motivated by the $248$~keV LUX--ZEPLIN candidate, we study pseudo-Dirac
dark matter coupled to a $0.20$~GeV dark photon with kinetic mixing
$\epsilon=8\times10^{-5}$. We consider masses of $45$ and $70$~GeV
with splittings of $0.80$ and $0.70$~MeV, respectively. We normalize each benchmark to one expected recoil in the LZ window. Both satisfy the dark-photon
search limits and the INTEGRAL/SPI limits on $\chi_H\to\chi_L+3\gamma$.
The recoil normalization fixes the coupling product $\alpha_D\epsilon^2$
that controls freeze-in from Standard Model fermions. We follow production
and entropy dilution during reheating, including pair production by
transverse and longitudinal dark photons. For $T_{\max}=m_\chi$ and
thermal dark photons, quadratic constant-width reheating gives the
observed abundance at $T_{\rm rh}\simeq3.0$ and $7.0$~GeV, compared
with $5.7$ and $11.7$~GeV for instantaneous reheating. At $m_\chi=45$~GeV, resonant $Z$ exchange raises the yield by a
factor of $18$ for kination ending at $T_{\rm rh}=5.96$~MeV, our
adopted lower limit, giving nearly twice the observed abundance. Reducing
the expected LZ count to $0.51$, well within the Poisson uncertainty of
one event, removes this excess. At $70$~GeV, on-shell $Z$ decay is closed
and kination gives the observed abundance at $T_{\rm rh}\simeq23$--$33$~MeV
as dark-photon fusion is varied from its thermal value to zero. Along the line of splittings giving a zero-speed xenon recoil of
$248$~keV, one expected LZ recoil and thermal dark photons give
a kination solution above $5.96$~MeV only for $m_\chi\gtrsim47$~GeV. A direct-detection signal of inelastic dark matter
can therefore constrain the expansion history before nucleosynthesis.}

\begin{document}
\maketitle

\section{Introduction}
\label{sec:intro}
The LUX--ZEPLIN (LZ) extended nuclear-recoil analysis reported one event at
$E_R=248\pm23\pm23$~keV in a $2.84$~tonne-year exposure, with a global
significance of $2.6\sigma$~\cite{LZ:2026axp}. One event does not establish a
dark-matter signal, but its high recoil energy and the absence of companion
events motivate inelastic scattering. Both the spectral shape and empty
sidebands enter this interpretation
\cite{Dent:2026bji,Rodd:2026tyn,Palmisano:2026kuj,Oliveira:2026jtn}.
Endothermic electroweak interpretations include Higgsinos,
mixing-suppressed singlet--doublet fermions and sneutrinos
\cite{Freese:2026sga,Wu:2026nhi,Lee:2026jxl,Lian:2026hpm}.
Exothermic vector-portal models provide another possibility
\cite{deLima:2026shq,Baer:2026fpy,Gemmell:2026yaw}.
In the exothermic case, a long-lived excited state releases energy when it
scatters. The recoil energy therefore need not come from the incident
particle's kinetic energy.

The same interaction must be consistent with the dark-matter abundance.
Freeze-in~\cite{Hall:2009bx,Bernal:2017kxu} can accommodate larger couplings
when the bath remains below the dark-matter mass, because pair production
then proceeds through its Boltzmann tail. This regime has been studied for
ultraviolet portals~\cite{Bernal:2025fcl} and electroweak
doublet fermions~\cite{Bernal:2026clv}.
Existing LZ studies consider low-temperature freeze-in
\cite{deLima:2026shq,Cabo-Almeida:2026uqw,He:2026idw,Unwin:2026qep},
modified expansion~\cite{Barman:2026omh}, and production followed by entropy
dilution~\cite{Langhoff:2026ujr,Pallis:2026jdc,Hor:2026voa}.
During a finite reheating phase, production also occurs above the
completion temperature. The resulting particles are diluted by later
entropy release~\cite{Chung:1998rq,Giudice:2000ex,Bernal:2024yhu,
Bernal:2024ndy,Bernal:2025fdr}. Boltzmann-suppressed
electroweak freeze-in has also been studied with finite reheating and
different equations of state before radiation domination
\cite{Bernal:2026vbg}. A radiation-dominated integral
starting at $\Trh$ can consequently miss most of the surviving abundance.

An exothermic vector portal also relates recoils to the decay
$\chi_H\to\chi_L+3\gamma$ below the electron-pair threshold
\cite{Krnjaic:2025zjl,McDermott:2017qcg}. Related photon constraints have been
studied with NuSTAR~\cite{Jeesun:2026ryo} and for transition charge-radius
and anapole interactions~\cite{Jodlowski:2026wjx}.
At fixed particle masses, the recoil rate and the present-day decay flux
depend on the same coupling product, so the photon limits apply whatever
the thermal history.

We study two exothermic benchmarks with $(m_\chi,\delta)=(45~{\rm GeV},
0.80~{\rm MeV})$ and $(70~{\rm GeV},0.70~{\rm MeV})$.
They share a $0.20$~GeV mediator and mixing $\epsilon=8\times10^{-5}$.
Their splittings are below $2m_e$, and their mediator lies in a directly
searched mass interval outside resonance vetoes. The two points differ in
an important production channel: on-shell $Z\to\chi_L\chi_H$ is open
at $45$~GeV and closed at $70$~GeV.
We first calculate their recoil spectra and normalizations, then examine
mediator searches and diffuse photons. We use the resulting couplings in
several finite reheating histories. A calculation with only dark-photon
exchange in the fermion amplitude isolates the effect of the $Z$
contribution. We also quantify mediator fusion and check chemical
equilibration, reannihilation and depletion of the excited state.

Each reheating history is specified by an equation of state and a cooling
exponent. The cooling law is written so that the bath entropy does not
decrease when $g_s$ changes through the QCD crossover. The standard
inflaton-decay scenarios correspond to particular values of the two
parameters.

\section{Particle model and recoil benchmarks}
\label{sec:model}
\subsection{Mass splitting and interactions}
Consider two Weyl fermions of opposite dark charge with Dirac mass $m_D$.
A scalar $S$ of charge two can generate equal Majorana masses $m_M$ after
$\langle S\rangle=v_D/\sqrt2$. For real equal Yukawa couplings $y$,
\begin{equation}
 {\cal M}_\chi=\begin{pmatrix}m_M&m_D\\m_D&m_M\end{pmatrix},\qquad
 m_M=\frac{yv_D}{\sqrt2},\qquad
 m_{H,L}\simeq m_D\pm m_M,\qquad \delta=2m_M.
 \label{eq:uv}
\end{equation}
The vector current is off diagonal in the Majorana basis. Small $y$ restores
a fermion-number symmetry and permits a splitting much smaller than $m_D$
\cite{Tucker-Smith:2001myb,Heeba:2023bik,Krnjaic:2025zjl}.
Alternative electroweak constructions relate the inelastic splitting to
anomaly-mediated supersymmetry breaking~\cite{Lee:2026tro}.
The splitting is independent of the dark-photon mass
$m_{\Aprime}=2g_Dv_D$.
For the two benchmarks below, $(v_D,y)$ is approximately
$(9.7~{\rm TeV},5.8\times10^{-8})$ and
$(6.3~{\rm TeV},7.8\times10^{-8})$.
Writing $S=(v_D+\rho)e^{iG/v_D}/\sqrt{2}$, the radial field $\rho$ is
the dark Higgs boson and $G$ is the Goldstone mode eaten by the dark
photon. We neglect the Higgs-portal coupling and take $m_\rho$ well
above the energies relevant for dark-matter pair production. In this
limit, expanding the Majorana mass term to second order in $G$ gives
the contact interaction used in Sec.~\ref{sec:source} to calculate
pair production by two longitudinal dark photons.

Hypercharge kinetic mixing gives
\begin{equation}
 {\cal L}\supset-\frac{\epsilon_Y}{2}B_{\mu\nu}F'^{\mu\nu}
 +ig_D\Aprime_\mu\bar\chi_L\gamma^\mu\chi_H,\qquad
 \alpha_D=\frac{g_D^2}{4\pi},\qquad \epsilon=\epsilon_Y\cos\theta_W.
 \label{eq:lagrangian}
\end{equation}
At low momentum transfer the light mediator couples to the electromagnetic
current with strength $\epsilon e$, up to corrections of order
$m_{\Aprime}^2/m_Z^2$~\cite{Holdom:1985ag,Pospelov:2008zw,Heeba:2023bik}.
The same mixing induces a $Z\chi_L\chi_H$ coupling of magnitude
\begin{equation}
 g_D\epsilon\tan\theta_W\frac{m_Z^2}{m_Z^2-m_{\Aprime}^2}.
 \label{eq:zchi}
\end{equation}
The production amplitude contains $\Aprime$ and $Z$ exchange.
The $Z$ pole contributes when $m_L+m_H<m_Z$. This condition separates
the two benchmarks introduced below.

\subsection{Down-scattering and the recoil likelihood}
For $\chi_H N\to\chi_L N$ with positive $\delta=m_H-m_L$,
\begin{equation}
 v_{\min}(E_R)=\frac{1}{\sqrt{2m_NE_R}}
 \left|\frac{m_NE_R}{\mu_{\chi N}}-\delta\right|,\qquad
 E_0=\frac{\mu_{\chi N}}{m_N}\delta.
 \label{eq:vmin}
\end{equation}
This expression applies to down-scattering~\cite{Graham:2010ca}.
At $E_R=E_0$, the incident particle can be at rest. Recoils away from this
energy require a nonzero speed to satisfy energy and momentum conservation.
Thus down-scattering has no overall incident-speed threshold, but has a
recoil-dependent $v_{\min}$. For a given speed $v$, its recoil interval is
\begin{equation}
 E_\pm(v)=\frac{\mu_{\chi N}^2}{2m_N}
 \left(\sqrt{v^2+\frac{2\delta}{\mu_{\chi N}}}\pm v\right)^2.
 \label{eq:support}
\end{equation}

The differential cross section is
\begin{equation}
 \frac{\dd\sigma_N}{\dd E_R}
 =\frac{8\pi\alpha_{\rm em}\alpha_D\epsilon^2Z^2m_N}
 {v^2(2m_NE_R+m_{\Aprime}^2)^2}F_N^2(E_R),
 \label{eq:dsigma}
\end{equation}
where $Z=54$ is the atomic number and $F_N$ is the Helm form
factor~\cite{Helm:1956zz,Lewin:1995rx},
\begin{equation}
 F_N(q)=\frac{3j_1(qR_1)}{qR_1}\,e^{-q^2s^2/2},\qquad
 R_1^2=c^2+\tfrac73\pi^2a^2-5s^2,\qquad q=\sqrt{2m_NE_R},
 \label{eq:helm}
\end{equation}
with $c=(1.23A^{1/3}-0.6)$~fm, $a=0.52$~fm and $s=0.9$~fm.
The rate per unit target mass is
\begin{equation}
 \frac{\dd R}{\dd E_R}=C_{\rm LZ}\,
 \frac{8\pi\alpha_{\rm em}Z^2\rho_\chi}{m_\chi}
 \sum_i\xi_i\,
 \frac{F_{N_i}^2(E_R)}{(2m_{N_i}E_R+m_{\Aprime}^2)^2}\,
 \eta\bigl(v_{\min,i}(E_R)\bigr),
 \label{eq:rate}
\end{equation}
\begin{equation}
 C_{\rm LZ}=\fH\alpha_D\epsilon^2,\qquad
 \fH=\frac{n_H}{n_L+n_H},\qquad
 \eta(v_{\min})=\int_{v>v_{\min}}\dd^3v\,\frac{f_E(\mathbf v)}{v},
 \label{eq:C}
\end{equation}
where the sum runs over the xenon isotopes with natural mass fractions
$\xi_i$, and $f_E$ is the halo velocity distribution in the Earth frame.
We use a truncated Maxwellian with $v_0=238$, $v_{\rm esc}=544$ and
$v_E=254$~km\,s$^{-1}$, and $\rho_\chi=0.3$~GeV\,cm$^{-3}$
\cite{Baxter:2021pqo,Freese:2012xd}.

At $E_R=248$~keV the momentum transfer is $q\simeq0.25$~GeV, comparable
to $m_{\Aprime}$, so the propagator in Eq.~\eqref{eq:rate} cannot be
replaced by a contact interaction. This momentum transfer also lies
beyond the first zero of the form factor.
For $^{131}$Xe the zeros of Eq.~\eqref{eq:helm} are at $E_R\simeq95$ and
$280$~keV, and the signal occupies the lobe between them. The falling
form factor, the propagator and $\eta$ together place the spectral
maximum below $E_0$, near $175$~keV for A and $182$~keV for B.
The rate on this lobe is sensitive to the nuclear radius: changing $R_1$
by $\pm3\%$ changes the expected count in the signal window by about
$\mp20\%$. The Helm parametrization is itself less reliable here than at
low momentum transfer~\cite{Ahmed:2026kan}.

\begin{figure}[t]
\centering\includegraphics[width=.9\textwidth]{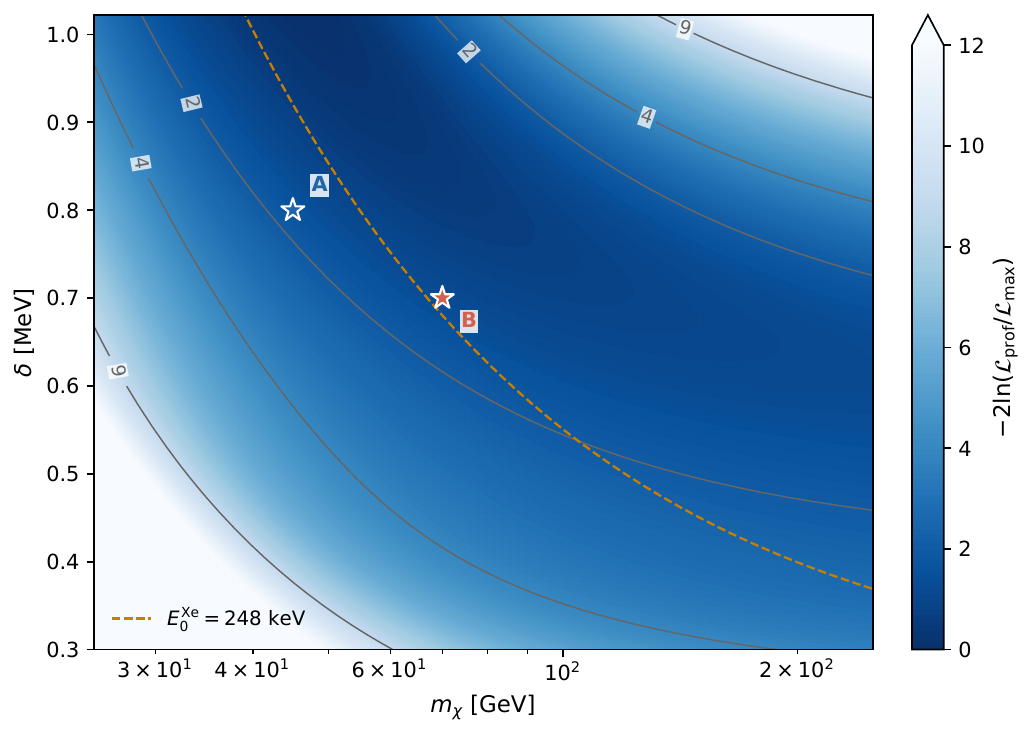}
\caption{Profiled phenomenological recoil likelihood at
$m_{\Aprime}=0.20$~GeV. The coupling normalization is fitted independently
at each point. Numbers on the contours denote
$-2\ln({\cal L}_{\rm prof}/{\cal L}_{\max})$.
The dashed line has $E_0^{\rm Xe}=248$~keV for a representative xenon
mass $122$~GeV. The stars mark benchmarks A and B. The scan includes the form
factor, propagator, isotope sum, energy kernel and empty sideband.}
\label{fig:recoil-plane}
\end{figure}

We use the phenomenological one-event likelihood of Ref.~\cite{deLima:2026shq}.
It has a true-recoil signal window $W=[5.4,270]$~keV, whose edges are the
energies at which the LZ nuclear-recoil efficiency falls to $50\%$, and an
empty high-energy interval $B=[350,650]$~keV representing the sideband.
The observed event enters through a Gaussian kernel $G$ centered on
$248$~keV with width $32.5$~keV, equal to the quadrature sum of the quoted
statistical and systematic uncertainties. Let $r(E)$ be the expected
number of events per unit recoil energy in the $2.84$~tonne-year exposure
for $C_{\rm LZ}=1$, that is, Eq.~\eqref{eq:rate} multiplied by the
exposure and divided by $C_{\rm LZ}$. The extended likelihood for one
event in $W$ and none in $B$ is
\begin{equation}
 {\cal L}\propto C_{\rm LZ} I_E
 \exp[-C_{\rm LZ}(I_W+I_B)],\qquad
 I_X=\int_X\dd E\,r(E),\qquad
 I_E=\int_W\dd E\,r(E)G(E;248,32.5).
 \label{eq:likelihood}
\end{equation}
The exponential is the Poisson factor for the expected total count, and
$C_{\rm LZ}I_E$ is the event density at the reconstructed energy.
Maximizing over $C_{\rm LZ}$ gives $\widehat C=1/(I_W+I_B)$, one expected
event in $W\cup B$, and a profiled likelihood proportional to
$I_E/(I_W+I_B)$. This ratio is the probability density for a single
event drawn from the spectrum to be reconstructed at $248$~keV. It
depends only on the spectral shape and is shown in
Fig.~\ref{fig:recoil-plane}.

The likelihood contains no background term, so the event is attributed
to the signal. Recoils between $270$ and $350$~keV belong to neither
interval.
The expected counts there are $0.004$ for A and $0.026$ for B, so this
choice does not affect the benchmarks. LZ's own likelihood is built in
scintillation and ionization observables~\cite{LZ:2026axp}, and $B$ is an
approximate true-energy representation of its sideband.

We take the common mediator parameters and excited fraction to be
\begin{equation}
 m_{\Aprime}=0.20~{\rm GeV},\qquad
 \epsilon=8.0\times10^{-5},\qquad f_H=\frac12.
 \label{eq:benchmark}
\end{equation}
The value $f_H=1/2$ is the outcome of freeze-in production, as shown in
Sec.~\ref{sec:states}.
Table~\ref{tab:benchmarks} defines benchmarks A and B.
The dark coupling at each point is fixed by one expected true recoil in
$W$ for the $2.84$~tonne-year exposure. Both points have a small expected
count in $B$. The maximum-likelihood amplitudes at fixed masses have
$N_W=0.999$ and $0.978$, respectively, so setting $N_W=1$ has negligible
normalization cost. The likelihood in Fig.~\ref{fig:recoil-plane} is
maximal at $m_\chi\simeq48$~GeV on the upper edge of the scan,
$\delta=2m_e$, above which $\chi_H$ decays to an electron pair. Relative
to this maximum, A and B have
$-2\Delta\ln\mathcal L\simeq1.66$ and $1.27$. Both lie inside the $-2\Delta\ln\mathcal L=2.30$ contour.
We use the customary two-parameter likelihood levels to compare shapes;
their coverage has not been calibrated for one event. We choose them
for their position relative to the $Z$ threshold: $2m_\chi=90$~GeV at A
and $140$~GeV at B.

\begin{table}[t]
\centering
\caption{Particle benchmarks at the common parameters in
Eq.~\eqref{eq:benchmark}. The normalization gives $N_W=1$ before
acceptance, and $N_B$ is the corresponding expected count in $B$.
The $Z$ channel refers to the on-shell two-body decay.}
\label{tab:benchmarks}
\begin{tabular}{@{}lcc@{}}
\toprule
Quantity & A & B \\\midrule
$m_\chi$ [GeV] & $45$ & $70$\\
$\delta$ [MeV] & $0.80$ & $0.70$\\
$\alpha_D$ & $8.412\times10^{-12}$ & $1.985\times10^{-11}$\\
$\alpha_D\epsilon^2$ & $5.384\times10^{-20}$ & $1.270\times10^{-19}$\\
$C_{\rm LZ}$ & $2.692\times10^{-20}$ & $6.351\times10^{-20}$\\
$N_B$ & $8.99\times10^{-4}$ & $2.25\times10^{-2}$\\
$Z\to\chi_L\chi_H$ & Open & Closed\\\bottomrule
\end{tabular}
\end{table}

These are true-recoil counts before acceptance and detector response.
LZ reports an average NR efficiency of $96\%$ between $14$ and $250$~keV,
with $50\%$ efficiency near $5.4$ and $269.9$~keV~\cite{LZ:2026axp}.
The rates below use this true-recoil convention. A detector-response fit
would determine the accepted count and refit the couplings.
The normalization $N_W=1$ does not place the spectral maximum at the
event. The central $68\%$ of the expected window events lies at
$155$--$207$~keV for A and $159$--$214$~keV for B, and only $10\%$ and
$15\%$ of them fall above $215.5$~keV, within one kernel width of
$248$~keV. This is the origin of the likelihood penalty quoted above.
The normalization also carries the Poisson uncertainty of a single
event. Its central $68\%$ interval is $0.17<N_W<3.3$, and the coupling
$\alpha_D$ scales linearly with $N_W$.

\begin{figure}[t]
\centering\includegraphics[width=.88\textwidth]{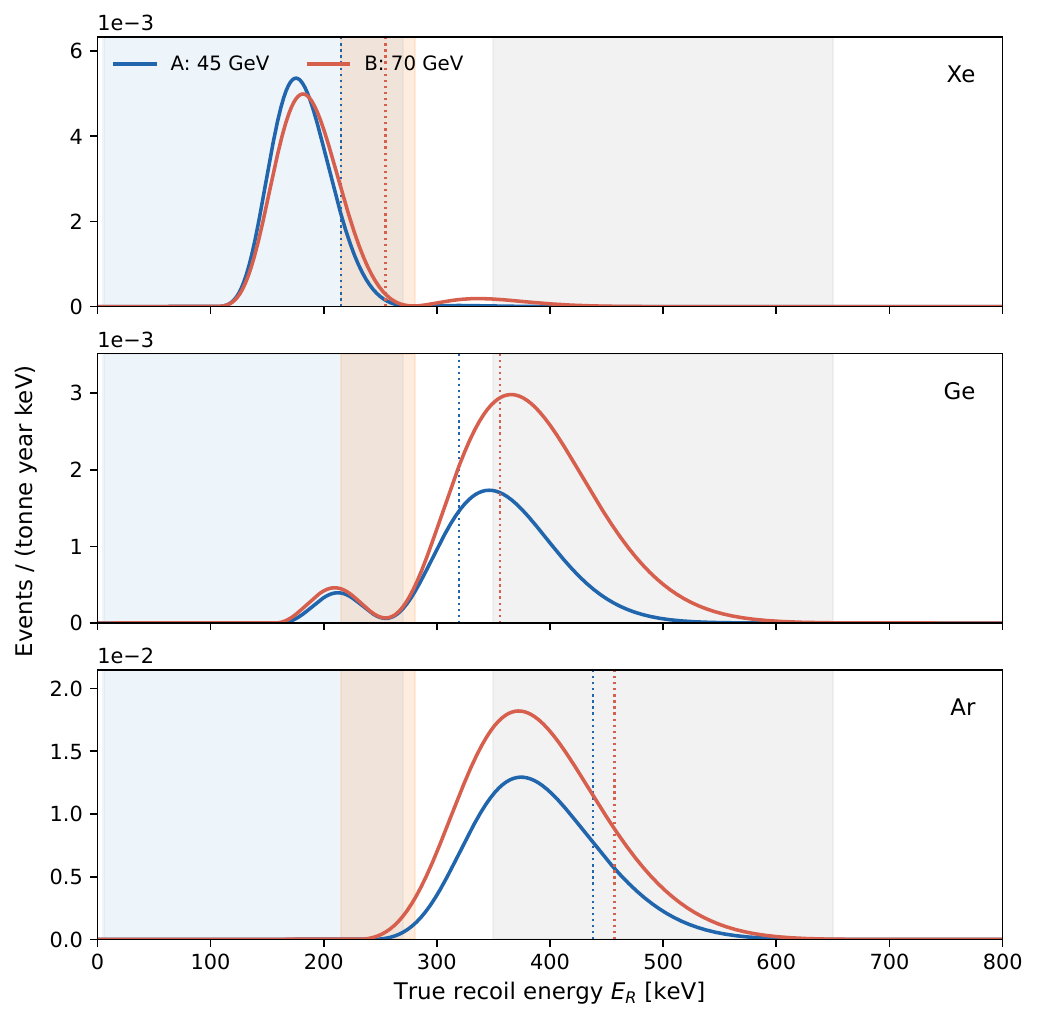}
\caption{Absolute true-recoil spectra for A (blue) and B (red), per
tonne-year and per keV. Dotted lines mark their isotope-averaged $E_0$.
Blue shading denotes $W$, grey shading denotes $B$, and orange shading
denotes $248\pm32.5$~keV. These intervals refer to the xenon likelihood.
The curves are evaluated before efficiency and energy reconstruction.}
\label{fig:targets}
\end{figure}

The isotope-averaged recoil centers $(E_0^{\rm Xe},E_0^{\rm Ge},E_0^{\rm Ar})$
are approximately $(215,319,438)$~keV for A and $(255,356,457)$~keV for B.
Figure~\ref{fig:targets} gives the absolute spectra. At the largest halo
speed, $798$~km\,s$^{-1}$, Eq.~\eqref{eq:support} gives about
$102$--$454$~keV for A and $102$--$638$~keV for B in $^{131}$Xe.
Conventional low-energy xenon windows consequently receive no signal.
The lighter targets give higher recoil energies. Argon has a larger integrated rate than xenon at both benchmarks. Integrated over
all recoil energies we find $0.23$ and $0.46$ events per tonne-year in
germanium and $1.8$ and $2.8$ events per tonne-year in argon for A and B,
almost all between $250$ and $600$~keV. For the $758$~tonne-day exposure of DEAP-3600~\cite{DEAP:2019yzn} this
corresponds to about $3.7$ and $5.8$ recoils before fiducial and
acceptance cuts,
at energies above its WIMP region of interest. Existing
argon data with an extended energy range could therefore test both
benchmarks~\cite{Baer:2026yrt,DarkSide-20k:2017zyg}. In xenon the
prediction is $0.35$ events per tonne-year between $5.4$ and $270$~keV.

\section{Laboratory and photon constraints}
\label{sec:constraints}
\subsection{Visible mediator searches}
Since $m_{\Aprime}<m_L+m_H$, the mediator has no invisible decay into dark
matter, and searches for visibly decaying dark photons constrain
$(m_{\Aprime},\epsilon)$ directly. The tree-level electron width is
\begin{equation}
 \Gamma_{\Aprime}\simeq\Gamma_{e^+e^-}
 =\frac{\alpha_{\rm em}\epsilon^2m_{\Aprime}}{3}
 \left(1+\frac{2m_e^2}{m_{\Aprime}^2}\right)
 \sqrt{1-\frac{4m_e^2}{m_{\Aprime}^2}}
 \simeq3.11\times10^{-12}~{\rm GeV}.
 \label{eq:visiblewidth}
\end{equation}
Since $m_{\Aprime}<2m_\mu$, this is the total width up to negligible
$3\gamma$ and $\nu\bar\nu$ modes. Thus
$\tau_{\Aprime}\simeq2.11\times10^{-13}$~s and
$c\tau_{\Aprime}\simeq63~\mu$m. The mediator decays promptly on collider
scales and long before reaching a beam-dump decay volume.

BABAR directly searched this mass in $e^+e^-\to\gamma\Aprime$ followed by
visible dileptons. Its reported mixing limits lie at
$\epsilon\sim10^{-4}$--$10^{-3}$~\cite{BaBar:2014zli}.
Our $\epsilon=8\times10^{-5}$ is below that range, and the mediator mass
is outside its resonance vetoes. APEX also searched $175$--$250$~MeV,
with sensitivity near $\epsilon^2\sim10^{-6}$~\cite{APEX:2011dww}.
KLOE's $\phi\to\eta\Aprime$ search constrains
$\epsilon^2\lesssim8\times10^{-6}$ over $50$--$210$~MeV
\cite{KLOE-2:2012lii}. KLOE radiative dielectron production reaches $\epsilon^2\sim10^{-6}$
in the $5$--$520$~MeV range~\cite{Anastasi:2015qla}. The A1/MAMI search
reaches $8\times10^{-7}$ over $40$--$300$~MeV~\cite{Merkel:2014avp}.
These limits are all weaker than the benchmark value
$\epsilon^2=6.4\times10^{-9}$.

Several searches require a different mass or decay length.
The published HPS result covers prompt masses up to $179$~MeV, while its
canonical displaced search covers $60$--$150$~MeV~\cite{Adrian:2022nkt}.
The 2026 FASER result reports exclusions extending to approximately
$150$~MeV~\cite{FASER2026}. The NA62 beam-dump analysis excludes
$10^{-6}\lesssim\epsilon\lesssim4\times10^{-5}$ for masses between $50$
and $600$~MeV~\cite{NA62:2023nhs}. Its upper edge in $\epsilon$ is set
by decay before the decay volume, and it does not reach the benchmark
mass at $\epsilon=8\times10^{-5}$. At these parameters, even at $400$~GeV energy, mediator has a
mean decay length is below $0.13$~m, far shorter than the roughly
$80$~m to the NA62 decay volume. Older downstream beam-dump searches
likewise lose the signal through decay before the fiducial volume.
The $\pi^0\to\gamma\Aprime$ channel used by NA48/2 is closed
\cite{NA482:2015wmo}.

Dimuon searches do not reach this mass. LHCb requires
$m_{\Aprime}>2m_\mu$~\cite{LHCb:2019vmc}, CMS starts at
$1.1$~GeV~\cite{CMS:2023hwl}, and the combined KLOE analysis covers
$519$--$987$~MeV~\cite{KLOE-2:2018kqf}.
Invisible-mediator searches at BABAR and NA64 assume an invisible decay
branch that is absent here. Dark-pair thresholds of $90$ and $140$~GeV
are also beyond the $B$ factories. Electroweak precision observables
constrain much larger mixings. At A, the induced invisible $Z$ width is
\begin{equation}
 \Gamma_{Z\to\chi_L\chi_H}\simeq1.18\times10^{-19}~{\rm GeV},
 \label{eq:zwidth}
\end{equation}
far below the experimental precision~\cite{ParticleDataGroup:2024cfk}. The decay is closed
at B. We neglect the Higgs portal, so there is no direct Higgs--dark-fermion
Yukawa coupling in this model.
A light dark Higgs would introduce additional searches. We neglect Higgs-portal mixing and take the radial scalar to be decoupled, as in Sec.~\ref{sec:model}.

\begin{figure}[t]
\centering\includegraphics[width=.9\textwidth]{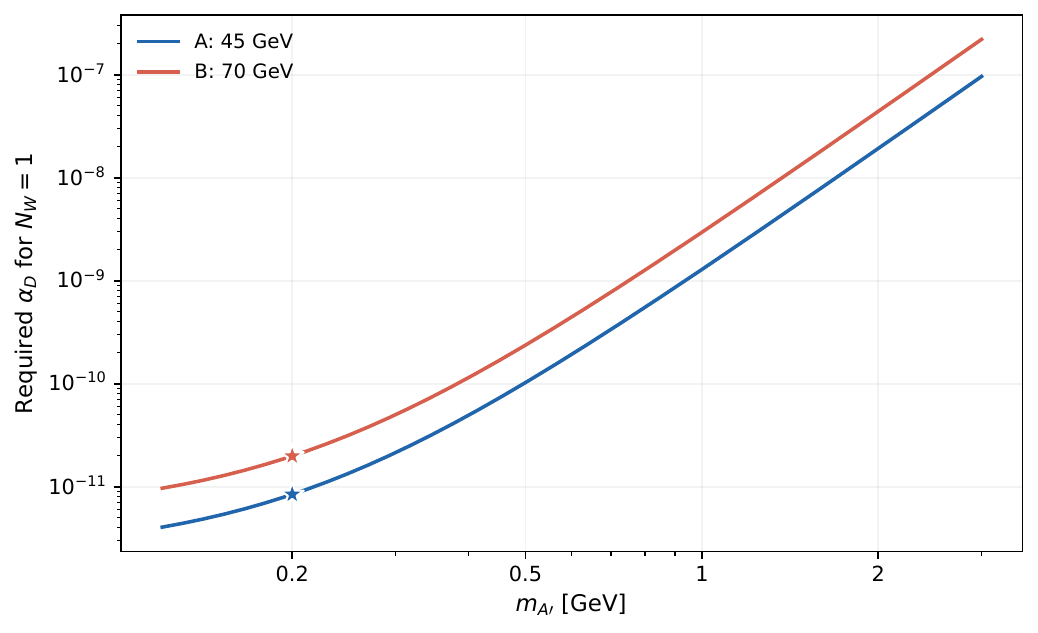}
\caption{Dark coupling required for one true recoil in $W$ as the
mediator mass varies, keeping $(m_\chi,\delta)$ fixed at A or B and
$\epsilon=8\times10^{-5}$, $f_H=1/2$. Stars mark $m_{\Aprime}=0.20$~GeV.
The full propagator is retained. The laboratory limits discussed in the
text are evaluated at the stars.}
\label{fig:mediator}
\end{figure}

The coupling required by the recoil rate depends on the mediator mass
through the propagator in Eq.~\eqref{eq:dsigma}, as shown in
Fig.~\ref{fig:mediator}. For $m_{\Aprime}$ well above the momentum
transfer, the required $\alpha_D\epsilon^2$ grows as $m_{\Aprime}^4$.
The mediator limits, in contrast, depend only on $(m_{\Aprime},\epsilon)$
as long as the decay to dark matter is closed. A point in the
$(m_\chi,\delta)$ plane must therefore be supplemented by the mediator
parameters before it can be compared with them.
Collider bounds have also been studied for elastic interpretations of the
LZ event using contact operators~\cite{Anwar:2026ffj}.
Those limits assume a mediator above the relevant momentum transfer.
The light mediator considered here must instead be retained explicitly
in the collider production amplitude.

\subsection{Three-photon decay}
\label{sec:photons}
Although the real $0.20$~GeV mediator decays visibly, the mediator in
$\chi_H$ decay is virtual and carries invariant mass at most
$\delta=0.80$ or $0.70$~MeV. It cannot produce an electron pair. An electron loop
with one dark vector and two photon vertices vanishes by charge conjugation
(Furry's theorem). The loop with three photons has four vector vertices
and survives. The leading photon channel is consequently
$\chi_H\to\chi_L\Aprime^*\to\chi_L+3\gamma$.
The neutrino channel is suppressed for a light kinetically mixed vector
\cite{Krnjaic:2025zjl}.

INTEGRAL/SPI observations also constrain three-photon decays of real
sub-MeV dark photons~\cite{Linden:2024fby}.
Related trident signals can arise from dark-matter annihilation into
long-lived mediators in celestial objects
\cite{Linden:2024uph}.
Those processes involve real mediators. The decay considered here
proceeds through a virtual mediator, with the splitting setting the
photon energies. We use the excited-state lifetime limits below.

To leading order in $\delta/m_\chi$ and $\delta/m_{\Aprime}$,
\begin{equation}
 \Gamma_{H\to L3\gamma}=
 \frac{g_D^2}{2\pi^2m_{\Aprime}^4}
 \int_0^{\delta^2}\frac{\dd q^2}{\sqrt{q^2}}
 (\delta^2-q^2)^{3/2}\Gamma_{\Aprime^*\to3\gamma}(\sqrt{q^2}).
 \label{eq:threephoton}
\end{equation}
We fold the full electron-loop enhancement tabulated in
Ref.~\cite{McDermott:2017qcg} into this integral. It increases the width by
$2.624$ at A and $1.990$ at B relative to the leading Euler--Heisenberg
result. The lifetimes are
\begin{equation}
 \begin{array}{c|cc}
 &\tau_H~[{\rm s}]&m_\chi\tau_H~[{\rm GeV\,s}]\\\hline
 A&4.36\times10^{23}&1.96\times10^{25}\\
 B&1.38\times10^{24}&9.67\times10^{25}
 \end{array}
 \label{eq:lifetime}
\end{equation}
The leading width scales as
$\alpha_D\epsilon^2\delta^{13}/m_{\Aprime}^4$.
Moving the splitting below $1$~MeV thus has a much larger effect than
splitting the same coupling product differently between $\alpha_D$ and
$\epsilon$.

The vector-mediator INTEGRAL/SPI limit in Fig.~5 of
Ref.~\cite{Krnjaic:2025zjl} is of order $10^{24}$~GeV\,s in this splitting
range for $f_H=1/2$. Both predictions in Eq.~\eqref{eq:lifetime} lie above it.
We use the published reference NFW halo, with local density
$0.3$~GeV\,cm$^{-3}$, and compare with its published lifetime curve.
At fixed masses, the decay flux scales as
$f_H\alpha_D\epsilon^2/m_\chi$. Changing $f_H$ while keeping the recoil
normalization fixed does not suppress it.

\subsection{Other dark-matter constraints}
The endothermic process $\chi_L N\to\chi_H N$ requires
$v\geq\sqrt{2\delta/\mu_{\chi N}}$. For xenon this is about
$2090$~km\,s$^{-1}$ for A and $1680$~km\,s$^{-1}$ for B, above the halo support.
The usual elastic spin-independent limits therefore cannot be applied to
the tree-level transition cross section. The XENON1T EFT analysis extends
to $54.4$~keV~\cite{XENON:2022avm}, and the first XENONnT NR analysis
extends to $60.5$~keV~\cite{XENON:2023cxc}. These energy windows lie below the xenon recoil support of A and B. A comparison with higher-energy data requires
the exothermic spectrum and that analysis's energy response, rather than
a rescaling of its elastic cross-section curve. A pair of transition vertices can
produce loop-induced elastic scattering, but its additional powers of
the portal products in Table~\ref{tab:benchmarks} make it negligible here
\cite{Heeba:2023bik}. Electron down-scattering releases an energy of order
$\delta$, rather than producing the low-energy spectrum assumed by light,
elastic dark-matter searches. Its much smaller reduced mass also suppresses
the rate relative to nuclear scattering.

Solar-neutrino searches provide a complementary test of inelastic
scattering~\cite{Pospelov:2026ewn,DiMauro:2026dqp,Nguyen:2026lui}.
Their sensitivity depends on capture, cooling, annihilation rates and
neutrino-producing final states. The state composition can change the
annihilation signal while leaving similar recoil spectra
\cite{Lian:2026bze}.
At our benchmarks, the small portal and dark couplings suppress
annihilation, while the light mediator decays predominantly to $e^+e^-$.

The threshold secluded-annihilation scale is
$\pi\alpha_D^2/m_\chi^2\lesssim2.6\times10^{-25}$~GeV$^{-2}$ for both
points, or $3.0\times10^{-42}$~cm$^3$\,s$^{-1}$. It is far below CMB limits.
Self-scattering is perturbative, with $\alpha_Dm_\chi/m_{\Aprime}\ll1$,
and is too weak to affect halo structure. Regeneration
$\chi_L\chi_L\to\chi_H\chi_H$ requires relative speeds of approximately
$3580$ and $2680$~km\,s$^{-1}$ for A and B, beyond the standard Galactic halo
\cite{Baxter:2021pqo,Freese:2012xd,OHare:2026nqi}.
At $m_{\Aprime}=0.20$~GeV the mediator is too heavy for stellar cooling,
and its short decay length prevents a free-streaming supernova energy
loss. Dark-matter production in such environments is kinematically
inaccessible. The mediator decays long before BBN. For $\chi_H\to\chi_L+3\gamma$, the fractional energy-injection rate
$f_H\delta/(m_\chi\tau_H)$ is $2.0\times10^{-29}~{\rm s}^{-1}$ at A
and $3.6\times10^{-30}~{\rm s}^{-1}$ at B, well below the CMB
sensitivity discussed in Ref.~\cite{Kim:2026euj}.

\section{Production during finite reheating}
\label{sec:reheating}
\subsection{Entropy-consistent background}
We describe the dominant component by $\rho_\phi\propto a^{-3(1+\omega)}$. Reheating starts at a maximum bath temperature $\Tmax$ and ends at the
completion temperature $\Trh$, below which the Universe is radiation
dominated. We treat $\Tmax$ as a free parameter, with $\Tmax=m_\chi$ as
the reference value, and neglect production before the bath reaches
$\Tmax$.
For a constant number of bath degrees
of freedom, define $T\propto a^{-\alpha}$ during reheating.
When $g_s(T)$ varies, we extend this prescription using
\begin{equation}
 {a(T)\over a_{\rm rh}}=
 \left({\Trh\over T}\right)^{1/\alpha}
 \left({g_s(\Trh)\over g_s(T)}\right)^{1/(3\alpha)}.
 \label{eq:cooling}
\end{equation}
Thus $Tg_s^{1/3}\propto a^{-\alpha}$ exactly and
\begin{equation}
 S\equiv sa^3\propto a^{3(1-\alpha)},\qquad
 J(T)\equiv1+\frac13\frac{\dd\ln g_s}{\dd\ln T},\qquad
 \dd t=-\frac{J(T)}{\alpha H(T)}\dd\ln T.
 \label{eq:entropy}
\end{equation}
For $\alpha\leq1$, the bath entropy is nondecreasing.
In contrast, applying a strict $T\propto a^{-\alpha}$ while separately
inserting a varying $g_s$ would give
$\dd\ln S/\dd\ln a=3-\alpha[3+\dd\ln g_s/\dd\ln T]$.
This can become negative through a rapid change in $g_s$.

With $\nu=3(1+\omega)/(2\alpha)$, our matching prescription is
\begin{equation}
 H(T)=H_{\rm rad}(\Trh)
 \left({T\over\Trh}\right)^\nu
 \left({g_s(T)\over g_s(\Trh)}\right)^{\nu/3},\qquad
 H_{\rm rad}(T)=\sqrt{\frac{\pi^2g_\rho(T)}{90}}\frac{T^2}{\mpl},
 \label{eq:H}
\end{equation}
where $\mpl=2.435\times10^{18}$~GeV. We require
$0<\alpha\leq\min[1,3(1+\omega)/4]$ and check $H(T)\geq H_{\rm rad}(T)$
throughout each numerical trajectory. The upper relation ensures
pre-completion dominance in the constant-$g$ limit.
The branch $\omega=1$, $\alpha=1$ has conserved visible entropy and
represents adiabatic kination.

Equations~\eqref{eq:cooling} and \eqref{eq:H} define a phenomenological
background with a sharp radiation-era match. They do not resolve the final
transition between the two dominant components. The choices
$(\omega,\alpha)=(0,3/8)$, $(1/3,1/4)$, $(1/3,1/2)$ and
$(1/3,3/4)$ reproduce the constant-$g$ asymptotes for quadratic
constant-width reheating and quartic reheating with scalar decay, constant
width and fermion decay, respectively
\cite{Giudice:2000ex,Bernal:2024yhu,Bernal:2025fdr,Deka:2026psw,
Bernal:2026vbg}.
We assume an equilibrated visible bath and no direct inflaton production
of dark matter. Production before visible thermalization can depend on
additional interactions~\cite{Harigaya:2019tzu,Cosme:2024ndc}.
We require $\Trh>5.96$~MeV~\cite{Barbieri:2025moq}. This bound was derived for
reheating by the decay of a massive particle, and we apply it to all
histories for uniformity. For kination, which injects no entropy, the
constraint arises from the expansion rate during nucleosynthesis and
need not coincide with it~\cite{Dutta:2010cu}. Section~\ref{sec:results} states how the
kination results depend on this choice.

\subsection{Collision terms}
\label{sec:source}
The leading Standard Model process is $f\bar f\to\chi_L\chi_H$.
For common dark mass $m_\chi$ and an effectively massless initial fermion,
\begin{equation}
 {\sigma_f(s)\over\alpha_D\epsilon^2}
 =\frac{4\pi\alpha_{\rm em}N_c^f}{3}
 (s+2m_\chi^2)\sqrt{1-\frac{4m_\chi^2}{s}}
 (|V_f|^2+|A_f|^2).
 \label{eq:sigmaff}
\end{equation}
This convention is spin averaged and color summed. Define
$D_A=(s-m_{\Aprime}^2)^{-1}$,
$D_Z=(s-m_Z^2+im_Z\Gamma_Z)^{-1}$ and
\begin{equation}
 D_N={m_{\Aprime}^2D_A-m_Z^2D_Z\over
 2c_W^2(m_Z^2-m_{\Aprime}^2)},\qquad
 V_f=Q_fD_A+(T_{3f}-2Q_fs_W^2)D_N,\qquad A_f=T_{3f}D_N.
 \label{eq:amplitudes}
\end{equation}
The relative sign gives the dark-photon--$Z$ interference. We use
$s_W^2=0.23122$, $m_Z=91.1876$~GeV and $\Gamma_Z=2.4952$~GeV.
We include $e,\mu,\tau,u,d,s,c,b$ and three active neutrinos.
Their physical masses are negligible compared with the dark-pair threshold.
The top channel is included with its threshold factors.

For the sum of these conventional cross sections, the total dark-number
source with Maxwell--Boltzmann initial distributions is
\begin{equation}
 {\cal C}_{f\bar f}^{\rm MB}(T)=
 {T\over4\pi^4}\sum_f\int_{4m_\chi^2}^\infty
 \dd s\,s^{3/2}\sigma_f(s)K_1(\sqrt s/T).
 \label{eq:Cff}
\end{equation}
The incoming spin sum and the two produced dark particles are included.
For chiral neutrinos the spin-summed amplitude yields the same convention.
We implement Fermi--Dirac distributions for the effectively massless
initial states. Electroweak $W^+W^-$ and $Zh$ production is included in
the Maxwell--Boltzmann approximation, with their initial polarization sums.
Together with the top channel, they contribute less than $2\%$ of the
final yield at both benchmarks. Appendix~\ref{app:source} gives the
normalizations.

All of these sources are proportional to $\alpha_D\epsilon^2$, the
combination that the recoil rate fixes through $C_{\rm LZ}$ at given
$f_H$. Once $N_W$ is chosen, the yield from Standard Model annihilation
depends on $m_\chi$, $\delta$ and $m_{\Aprime}$, but not on how the
product is divided between $\alpha_D$ and $\epsilon$. That division
enters only the mediator-fusion source below, which scales as
$\alpha_D^2$, and the laboratory limits on $\epsilon$.

A thermal dark-photon population can also produce pairs through
$\Aprime\Aprime\to\chi_L\chi_L,\chi_H\chi_H$.
The transverse contribution in the pseudo-Dirac limit is the Breit--Wheeler result for
two transverse vectors~\cite{Breit:1934zz},
\begin{equation}
 {\sigma_{TT}(s)\over\alpha_D^2}
 ={2\pi\over s}\left[(3-\beta_\chi^4)
 \ln{1+\beta_\chi\over1-\beta_\chi}
 -2\beta_\chi(2-\beta_\chi^2)\right],\qquad
 \beta_\chi=\sqrt{1-4m_\chi^2/s}.
 \label{eq:BW}
\end{equation}
For $\delta\ne0$ the transition current is not conserved. In the regime
$m_{\Aprime}\ll\sqrt{s}\ll m_\rho$, the amplitude for two longitudinal
dark photons to produce a dark-matter pair grows with energy
\cite{Appelquist:1987cf,Kim:2026euj}. Expanding the Majorana mass term to second order
in the Goldstone field $G$ gives a coupling of magnitude
$m_M/v_D^2=2g_D^2\delta/m_{\Aprime}^2$.
Summing both identical-pair final states, the effective cross section entering the source is
\begin{equation}
 \sigma_{AA}(s)=\sigma_{TT}(s)
 +{2\pi\alpha_D^2\delta^2\over m_{\Aprime}^4}\,\beta_\chi^3 .
 \label{eq:longfusion}
\end{equation}
Here the polarization sum is divided by four, as in the transverse
average in Eq.~\eqref{eq:BW}. The second term therefore carries one
longitudinal pair per four transverse combinations. We retain the leading
terms for $m_{\Aprime}\ll m_\chi$ and $\delta\ll m_\chi$.
The $\beta_\chi^3$ factor suppresses the longitudinal contribution at threshold.
The identical-initial-particle factor $1/2$ cancels the two produced
particles. The resulting MB total-number source is
\begin{equation}
 {\cal C}_{AA}^{\rm MB}
 ={T\over8\pi^4}\int_{4m_\chi^2}^\infty
 \dd s\,s^{3/2}\sigma_{AA}(s)K_1(\sqrt s/T).
 \label{eq:CAA}
\end{equation}
We use Bose--Einstein initial distributions in the equilibrium estimate.
This relativistic kernel avoids a threshold-annihilation approximation at
$T$ comparable to $m_\chi$.
As a detailed-balance check, its nonrelativistic limit is
${\cal C}_{AA}=\langle\sigma v\rangle_{\chi\bar\chi\to AA}
(n_\chi^{\rm eq})^2/2$, where $n_\chi^{\rm eq}$ counts all four Dirac
degrees of freedom and $\langle\sigma v\rangle\to\pi\alpha_D^2/m_\chi^2$.

Whether the mediator is thermally populated at the relevant temperatures
is not obvious. In the relativistic plasma the photon has a plasma
frequency $\omega_p\simeq0.26\,T$ for the charged species lighter than
the top quark, which exceeds $m_{\Aprime}$ for $T\gtrsim0.8$~GeV. Away from resonance, transverse production is suppressed by a factor of order
$m_{\Aprime}^4/\omega_p^4$. The longitudinal rate has a separate in-medium polarization function~\cite{Hambye:2019dwd}. Pair production from fermions is not
affected, since it proceeds at $s\geq4m_\chi^2\gg\omega_p^2$.
We therefore evaluate the mediator-fusion source in two limits, a thermal
population and none, and quote the resulting range of completion
temperatures. At the temperatures in Table~\ref{tab:histories}, dark-photon fusion
accounts for $0.04$--$0.67\%$ of the yield at A and $9$--$33\%$ at B. We assume no nonthermal source of mediators.

In the thermal limit the dark photon also contributes three polarizations
to $g_s$ and $g_\rho$. Removing this contribution lowers the completion
temperatures by less than $1\%$ for A and by $0.5\%$--$1.5\%$ for the
entropy-producing histories of B. The largest change is $5.4\%$, for
kination at B.

\subsection{Yield and entropy dilution}
With $N_\chi=a^3(n_L+n_H)$, dilute production gives
$\dot N_\chi=a^3{\cal C}$, where
${\cal C}={\cal C}_{\rm SM}+{\cal C}_{AA}$.
Taking $a_{\rm rh}=1$, Eqs.~\eqref{eq:cooling}--\eqref{eq:H} give
\begin{align}
 Y_{\rm reh}&={1\over s(\Trh)}\int_{\Trh}^{\Tmax}
 \dd\ln T\,{J(T)\over\alpha H(T)}a^3(T){\cal C}(T),\label{eq:Yreh}\\
 Y_{\rm RD}&=\int_0^{\Trh}\dd\ln T\,
 {J(T){\cal C}(T)\over s(T)H_{\rm rad}(T)}.
 \label{eq:Yrd}
\end{align}
The comoving-volume weight in $Y_{\rm reh}$ and its final entropy
normalization account for dilution. We solve
\begin{equation}
 Y_{\rm reh}+Y_{\rm RD}=Y_{\rm obs}
 ={\Omega_\chi h^2(\rho_c/h^2)\over m_\chi s_0}
 =\begin{cases}9.718\times10^{-12}&(A),\\
 6.248\times10^{-12}&(B).\end{cases}
 \label{eq:Yobs}
\end{equation}
using $\Omega_\chi h^2=0.12$, $s_0=2891.2$~cm$^{-3}$ and
$\rho_c/h^2=1.053672\times10^{-5}$~GeV\,cm$^{-3}$
\cite{Planck:2018vyg,ParticleDataGroup:2024cfk}.
Temperature-dependent $g_s$ and $g_\rho$ are interpolated from the
thermodynamic table specified in Appendix~\ref{app:numerics}.

\section{Excited fraction and freeze-in consistency}
\label{sec:states}
Fermion annihilation injects one particle in each state. The summed
mediator channels populate the two states equally in the small-splitting
limit. Let $f=N_H/N_\chi$. Including self-conversion, its evolution has
the form
\begin{equation}
 \dot f=(1/2-f){\dot N_\chi\over N_\chi}
 -\langle\sigma v\rangle_{HH\to LL}n_\chi
 \left[f^2-(1-f)^2e^{-2\delta/T_\chi}\right]
 -\Gamma_H f+\left.\dot f\right|_{\rm bath}.
 \label{eq:states}
\end{equation}
Here $T_\chi$ is the kinetic temperature of the dark matter, $\Gamma_H$
is the three-photon width of Sec.~\ref{sec:photons}, and the last term is
the conversion by scattering on the bath discussed below.
The injection term matters while production continues.
For the perturbative transition potential, the low-velocity conversion
cross section is~\cite{deLima:2026shq}
\begin{equation}
 \langle\sigma v\rangle_{HH\to LL}
 ={4\pi\alpha_D^2m_\chi\sqrt{2m_\chi\delta}
 \over(2m_\chi\delta+m_{\Aprime}^2)^2}.
 \label{eq:conversion}
\end{equation}
Nonperturbative corrections relevant for stronger dark interactions
\cite{Baer:2026fpy} are absent at our coupling.

A useful conservative self-conversion check omits both reverse conversion
and the positive injection term while retaining the actual accumulated
total density. It gives
\begin{equation}
 f_H\geq {1\over2+\tau_{HH}},\qquad
 \tau_{HH}=\int\dd t\,
 \langle\sigma v\rangle_{HH\to LL}n_\chi(t).
 \label{eq:bound}
\end{equation}
We construct $n_\chi(t)$ from the cumulative comoving production during
reheating and radiation domination. The numerical bound uses twice the
zero-speed cross section to cover finite incident speeds in the Born
kernel, as discussed in Appendix~\ref{app:source}. For the histories of
Table~\ref{tab:histories}, $1/2-f_H<5.6\times10^{-8}$ for A and
$2.3\times10^{-7}$ for B
from this channel. The fractions of excited particles decayed by the
present epoch are $1.0\times10^{-6}$ and $3.15\times10^{-7}$. These effects
are negligible at percent accuracy.

Bath scattering $\chi_H f\leftrightarrow\chi_L f$ also changes the fraction
\cite{Heeba:2023bik}. We estimate this effect in the heavy-DM limit, retaining both directions. For a static dark particle, dropping
momentum suppression and final-state blocking gives the electron
contact-rate envelope
\begin{equation}
 \Gamma_e(T)={16\alpha_{\rm em}(\alpha_D\epsilon^2)\over\pi m_{\Aprime}^4}
 \left[I_4T^5+2\delta I_3T^4+\delta^2 I_2T^3\right],
 \label{eq:bath}
\end{equation}
where $I_n=\int_0^\infty\dd x\,x^n/(e^x+1)$.
Here $\Gamma_e$ is the down-scattering rate per $\chi_H$ from thermal
electrons and positrons. To include the other charged species, we take
$\Gamma_{\rm down}=8\Gamma_e$, rounding up the charge-weighted sum
$\sum_f N_c^f Q_f^2=20/3$ for three charged leptons and five quarks.
We treat these species as massless throughout. We evolve
$\dot f=-\Gamma_{\rm down}[f-(1-f)e^{-\delta/T}]$ from $T=m_\chi/10$,
starting at the equilibrium value there, and omit the production term,
which drives $f$ toward $1/2$. This gives $f_H\simeq0.4982$ for A,
$0.4978$ for the entropy-producing histories of B, and $0.4987$ for
kination at B. Propagator suppression, fermion masses and Pauli blocking
all reduce the true rate, so the shift of the recoil rate from its
$f_H=1/2$ value is below one percent. A full treatment would follow the
dark-matter momentum distribution, which is not thermal.

We test the dilute-production approximation using
${\cal C}/(n_\chi^{\rm eq}H)$ and the reannihilation ratio
$n_\chi\langle\sigma v\rangle/H$. Their largest values for the histories
of Table~\ref{tab:histories} at $\Tmax=m_\chi$ are $1.82\times10^{-6}$ and
$5.88\times10^{-15}$. The latter uses the threshold cross section for annihilation into dark photons. The dark matter never approaches chemical
equilibrium, and annihilation of the produced population is negligible.

\section{Results}
\label{sec:results}
\subsection{Two production regimes}
At A, $m_L+m_H\simeq90$~GeV lies below $m_Z$. The finite-width propagator
in Eq.~\eqref{eq:amplitudes} includes resonant production through
$Z\to\chi_L\chi_H$. We therefore add no separate $Z$-decay source.
At B, the $140$~GeV threshold closes this channel, while off-shell
$Z$ exchange still contributes.

\begin{figure}[t]
\centering\includegraphics[width=.98\textwidth]{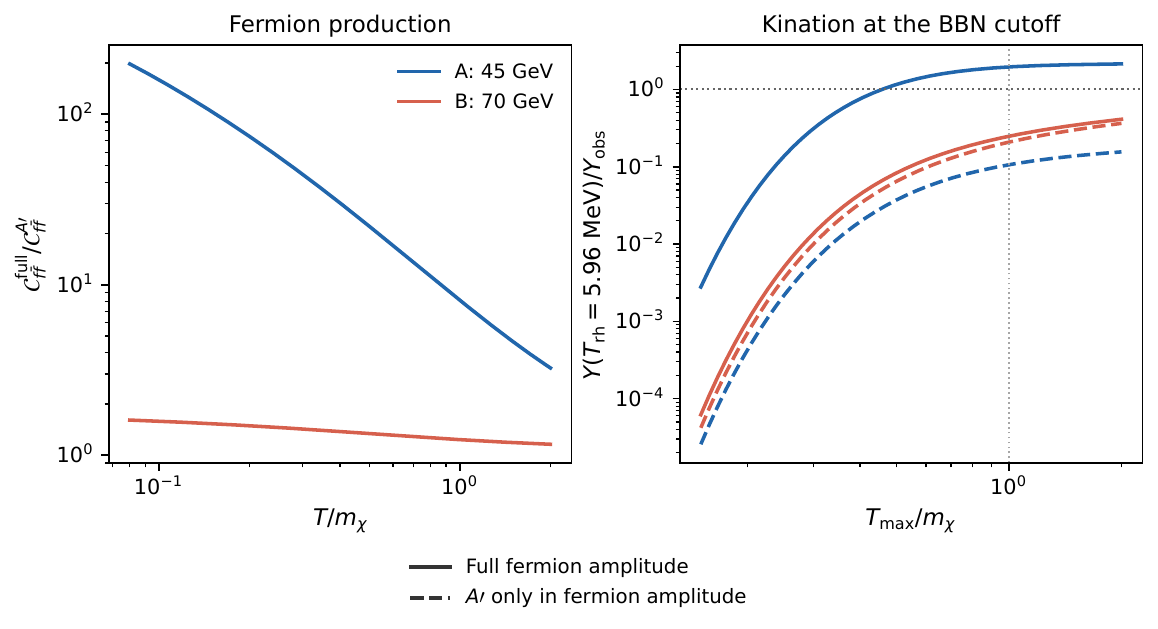}
\caption{Effect of $Z$ exchange. Left: full light-fermion source divided by its
$\Aprime$-only value for A and B. Right: kination yield at
$\Trh=5.96$~MeV versus $\Tmax/m_\chi$. Solid curves use the full fermion
amplitude. Dashed curves remove $Z$ exchange from that amplitude, with
all other channels and parameters unchanged. Values above the horizontal
$Y/Y_{\rm obs}=1$ line overproduce dark matter.}
\label{fig:z-production}
\end{figure}

The left panel of Fig.~\ref{fig:z-production} compares the light-fermion
source with the result obtained by retaining only $\Aprime$ exchange.
The resonance enhances production at A. The right panel shows its
consequence for kination, keeping the particle parameters and other
production channels fixed. At $\Tmax=m_\chi$ and
$\Trh=5.96$~MeV, the full calculation gives
\begin{equation}
 {Y_A\over Y_{{\rm obs},A}}=1.95,\qquad
 \left.{Y_A\over Y_{{\rm obs},A}}\right|_{\Aprime\ {\rm only\ in}\ f\bar f}
 =0.106.
 \label{eq:z-control}
\end{equation}
With the full amplitude, kination ending at $5.96$~MeV overproduces the
dark matter at A by a factor $1.95$, while the $\Aprime$-only amplitude
would allow a solution near $48.5$~MeV. Resonant $Z$ exchange thus raises
the kination yield at A by a factor of $18.4$, despite the extremely
small partial width in Eq.~\eqref{eq:zwidth}. At B the corresponding
factor is $1.19$.

At A, reducing the expected LZ count from $N_W=1$ to $0.51$ gives the
observed abundance at $\Trh=5.96$~MeV. At fixed mixing, the fermion
contribution scales as $N_W$ and the fusion contribution as $N_W^2$.
The value $N_W=0.51$ lies within the one-count Poisson interval
$0.17<N_W<3.3$ and gives $-2\Delta\ln{\cal L}=0.36$ for the signal
window alone. A $3\%$ change in the nuclear radius also shifts the
inferred coupling by about $20\%$.

Keeping $N_W=1$ instead gives $\Trh\simeq3.05$~MeV if we hold the
equation of state fixed below $4$~MeV. Whether nucleosynthesis allows
this temperature depends on the expansion rate during kination
(Sec.~\ref{sec:reheating}).

\subsection{Completion temperatures}
Table~\ref{tab:histories} gives the completion temperatures at
$\Tmax=m_\chi$ for each benchmark. The ranges vary the mediator-fusion
source from its equilibrium estimate to zero. A smaller source requires
a larger completion temperature. This variation is negligible at the
stated precision for the three GeV-scale solutions of A. It has a larger
effect at B.

\begin{table}[t]
\centering
\caption{Required completion temperatures at $\Tmax=m_\chi$. Entries are
in GeV. Ranges vary mediator fusion from all three equilibrium polarizations to zero at fixed bath degrees of freedom.
``No solution'' means that the observed abundance cannot be obtained
for $\Trh>5.96$~MeV.}
\label{tab:histories}
\begin{tabular}{@{}lcccc@{}}
\toprule
History & $\omega$ & $\alpha$ & A & B\\\midrule
Quadratic, constant width & $0$ & $3/8$ & $3.02$ & $7.04$--$7.18$\\
Quartic, scalar decay & $1/3$ & $1/4$ & $4.59$ & $9.48$--$9.57$\\
Quartic, constant width & $1/3$ & $1/2$ & $1.98$ & $5.08$--$5.27$\\
Quartic, fermion decay & $1/3$ & $3/4$ & $0.0910$--$0.0914$ & $0.259$--$0.307$\\
Adiabatic kination & $1$ & $1$ & No solution & $0.0233$--$0.0331$\\\midrule
Instantaneous radiation era & \multicolumn{2}{c}{---} & $5.705$ & $11.67$--$11.77$\\\bottomrule
\end{tabular}
\end{table}

\begin{figure}[t]
\centering\includegraphics[width=.98\textwidth]{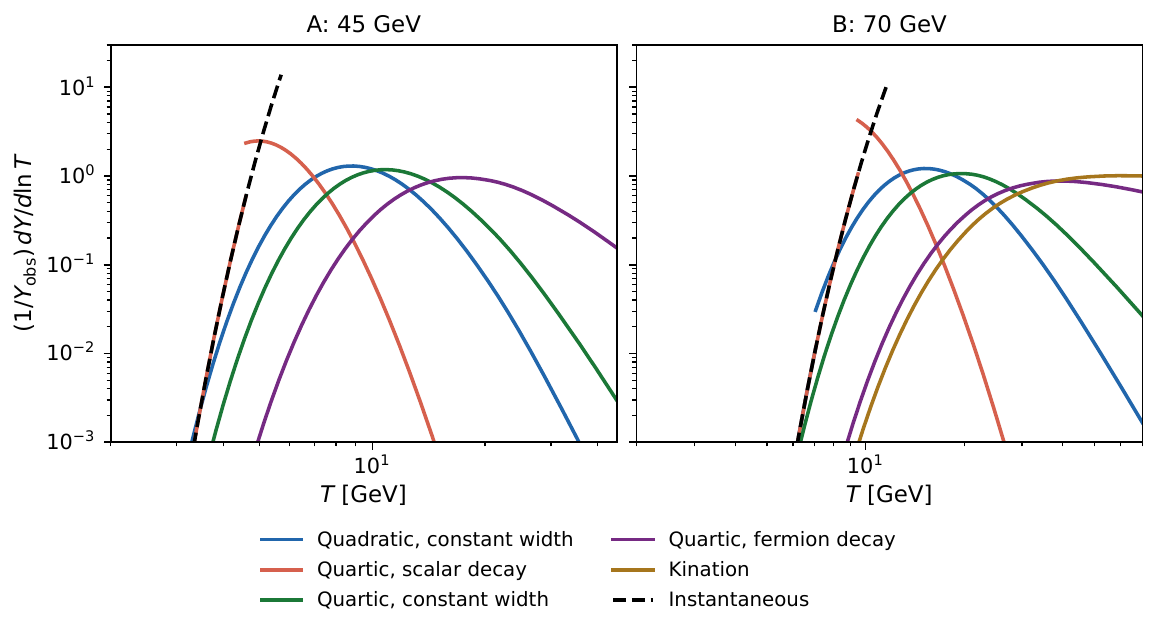}
\caption{Contribution to the final yield per logarithmic temperature
interval, normalized to each benchmark's $Y_{\rm obs}$, with the equilibrium
mediator source. The two panels use the temperatures in
Table~\ref{tab:histories}. The dashed curve denotes instantaneous
reheating. The reheating and radiation branches are plotted separately
at the change in the cooling law. Kination has no curve at A because
there is no solution at $\Tmax=m_\chi$.}
\label{fig:kernel}
\end{figure}

Figure~\ref{fig:kernel} shows the contribution to the final abundance
per logarithmic temperature interval. Each curve integrates to the
observed yield for its benchmark. The instantaneous calculation starts
at its radiation-era temperature. The finite histories include hotter
production and its subsequent entropy dilution.
The ordering of the completion temperatures in Table~\ref{tab:histories}
can be understood from Eqs.~\eqref{eq:Yreh} and \eqref{eq:Yrd}.
For approximately constant $g_s$ and $g_\rho$, entropy dilution gives
a factor $(\Trh/T)^{3/\alpha-3}$, while the expansion rate gives
$H_{\rm rad}(T)/H(T)=(\Trh/T)^{\nu-2}$. The contribution to the final
yield per logarithmic temperature interval during reheating is then
\begin{equation}
 \frac{\dd Y_{\rm reh}}{\dd\ln T}\simeq
 \frac{\mathcal C(T)}{\alpha\,s(T)H_{\rm rad}(T)}
 \left(\frac{\Trh}{T}\right)^p,\qquad
 p=\frac{3}{\alpha}-3+\nu-2
  =\frac{3(3+\omega)}{2\alpha}-5 .
 \label{eq:pexp}
\end{equation}
For the five finite histories in table order, $p=7,\,15,\,5,\,5/3$
and $1$. Larger $p$ suppresses the contribution from $T\gg\Trh$
more strongly. This is visible at benchmark B: the yield peaks near
$9.5$~GeV for quartic reheating through scalar decay ($p=15$),
near $15$~GeV for quadratic reheating ($p=7$), and near $61$~GeV
for kination ($p=1$).

Kination conserves visible entropy. Its faster expansion can reduce
production, but has no entropy release to dilute it. For A with
$\Tmax=m_\chi$, the yield at $\Trh=5.96$~MeV exceeds the observed
abundance by the factor $1.95$ of Eq.~\eqref{eq:z-control}. For B,
$\Trh\simeq23$~MeV gives the observed abundance.

\begin{figure}[t]
\centering\includegraphics[width=.98\textwidth]{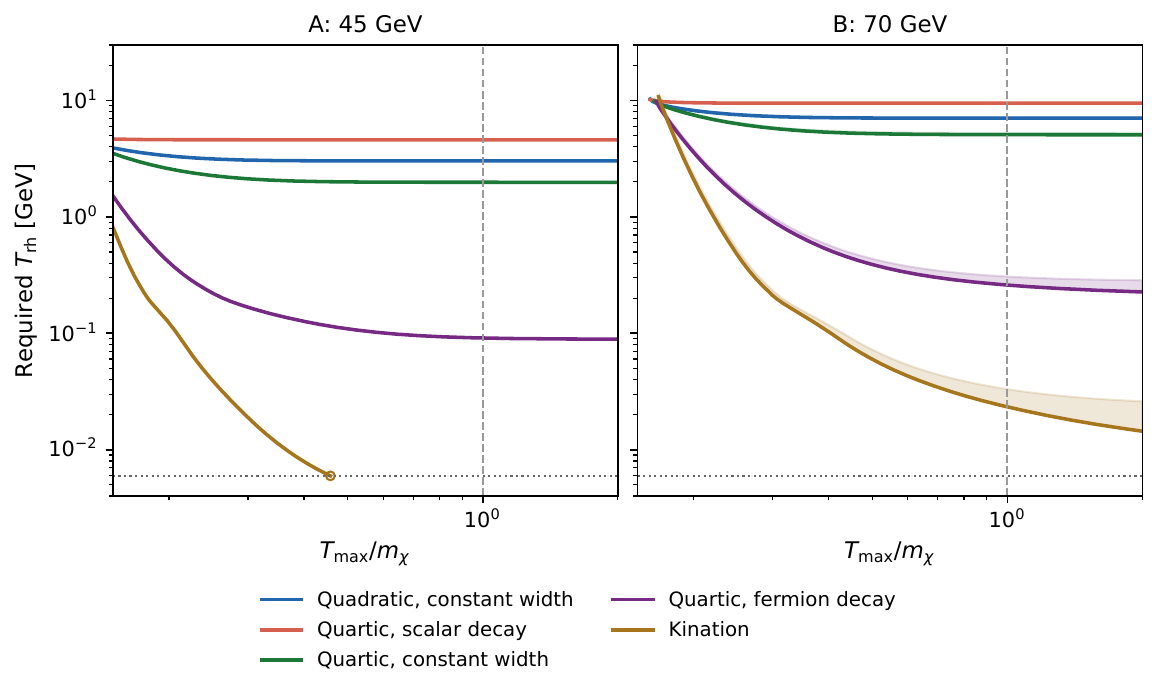}
\caption{Completion temperature versus maximum bath temperature for A
and B. Solid curves follow the smaller-$\Trh$ branch and include the equilibrium
mediator source. Shaded bands
extend to the zero-fusion result. The vertical line marks
$\Tmax=m_\chi$, and the horizontal line marks the BBN-motivated completion
cutoff. The open marker gives the kination endpoint at A, calculated by
solving $Y(\Trh=5.96~{\rm MeV})=Y_{\rm obs}$.}
\label{fig:trh}
\end{figure}

Figure~\ref{fig:trh} shows the dependence on $\Tmax$.
Raising $\Tmax$ opens an earlier production interval and generally lowers
$\Trh$. The entropy-producing histories become less sensitive once their
main production interval is covered. Kination retains more sensitivity
because it has no entropy dilution. At A it becomes viable for a
sufficiently low maximum temperature, before reaching the endpoint shown
in the figure. The scan extends to $\Tmax=2m_\chi$, or $90$ and $140$~GeV.
We use vacuum electroweak masses, couplings and widths over this range.

\begin{figure}[t]
\centering\includegraphics[width=.98\textwidth]{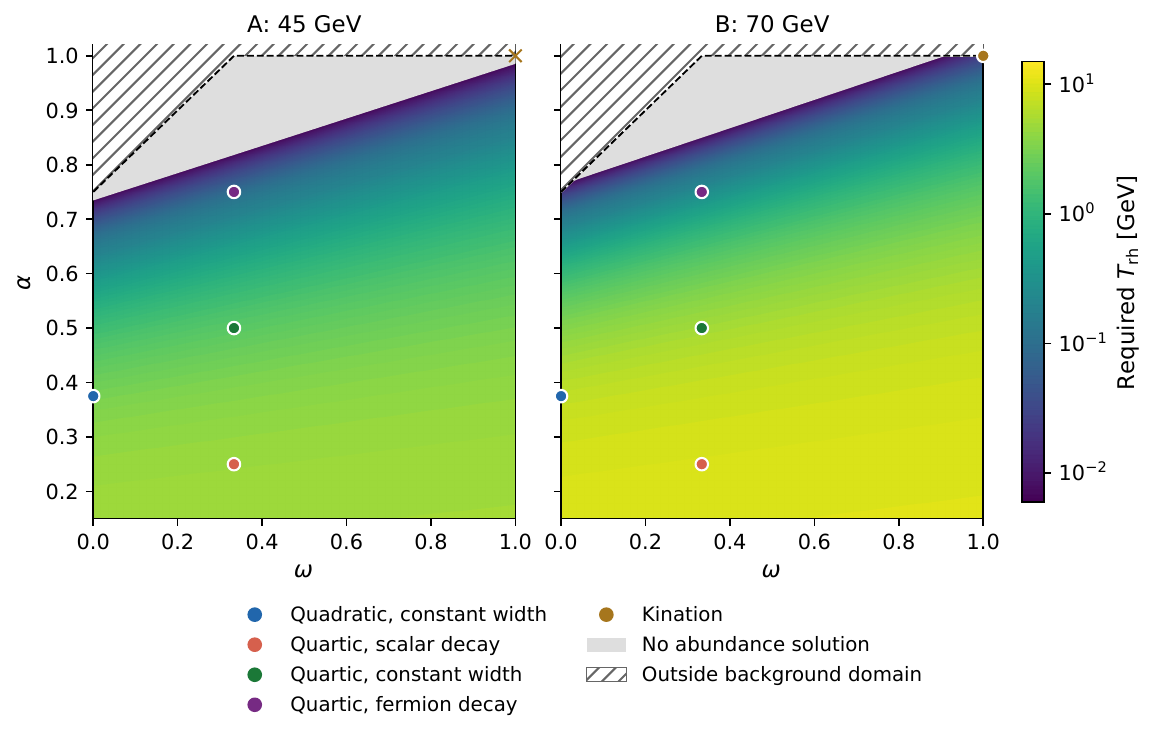}
\caption{Completion temperature at $\Tmax=m_\chi$ with the equilibrium
fusion estimate, using a common color scale. Grey denotes no abundance
solution above the BBN cutoff. Hatching denotes the background domain
$\alpha>3(1+\omega)/4$. Symbols locate the five histories of
Table~\ref{tab:histories}.
A cross marks the kination point without a solution at A.}
\label{fig:plane}
\end{figure}

The broader background dependence appears in Fig.~\ref{fig:plane}.
Increasing $\omega$ at fixed $\alpha$ speeds up expansion and usually
requires a larger completion temperature. Increasing $\alpha$ allows
more early production to survive and has the opposite effect.
The common color scale permits a direct temperature comparison between
the two particle points. We check bath-entropy growth, pre-completion
dominance, chemical equilibration and self-conversion throughout the grid.

\begin{figure}[t]
\centering\includegraphics[width=.9\textwidth]{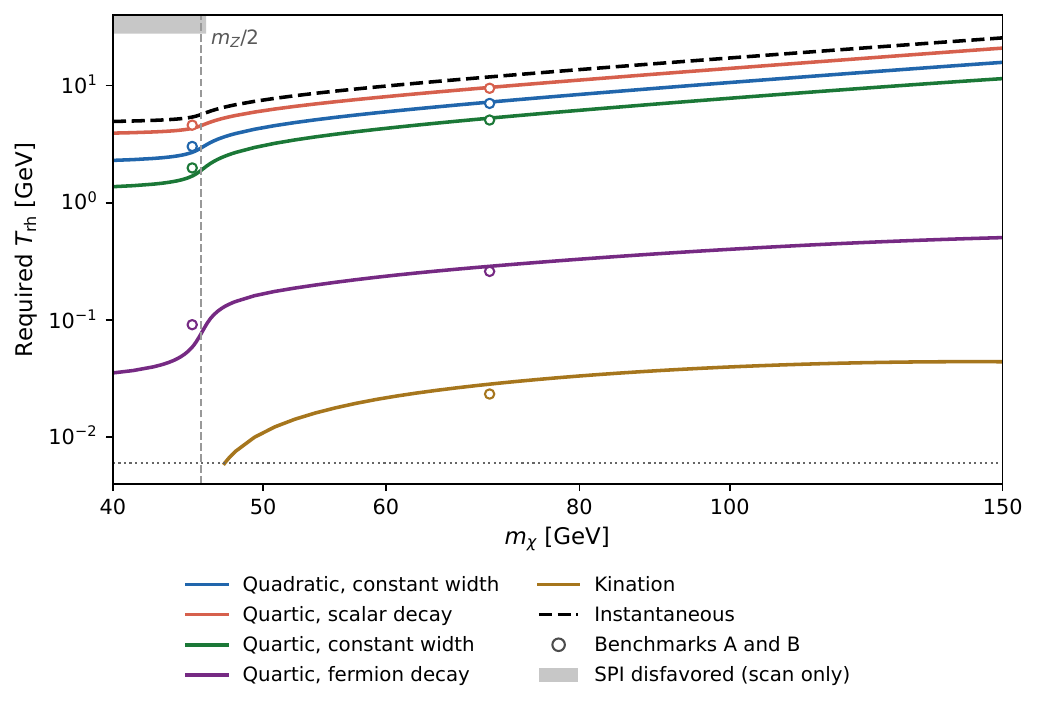}
\caption{Completion temperature at $\Tmax=m_\chi$ along the line
$E_0^{\rm Xe}=248$~keV, with $\alpha_D$ fixed by $N_W=1$ at each mass and
a thermal mediator population. The dashed curve is instantaneous
reheating. The vertical line marks $m_\chi=m_Z/2$ and the horizontal line
$\Trh=5.96$~MeV. Open circles show benchmarks A and B from
Table~\ref{tab:histories}. The grey strip at the top marks the scan points
disfavored by the approximate INTEGRAL/SPI limit of Sec.~\ref{sec:photons}
at $N_W=1$. Each benchmark is tested at its own splitting.}
\label{fig:mass-scan}
\end{figure}

The two benchmarks sample opposite sides of the $Z$ threshold.
Figure~\ref{fig:mass-scan} follows the completion temperature
continuously in $m_\chi$ along the dashed line of
Fig.~\ref{fig:recoil-plane}, $\delta=0.248~{\rm MeV}\,(m_\chi+122~{\rm
GeV})/m_\chi$, on which the xenon recoil center coincides with the event.
The scan starts at $40$~GeV, where $\delta=1.0044$~MeV is just below $2m_e$, and $\alpha_D$
is fixed by $N_W=1$ at each mass. Along this line $\alpha_D$ stays between
$1.7$ and $2.0\times10^{-11}$, so the mass dependence in the figure is
that of the production rate.

All histories show a step at $m_\chi\simeq m_Z/2$, smoothed by the $Z$
width. Between $43$ and $48$~GeV the completion temperature rises by a
factor $1.4$ for instantaneous reheating, $1.6$ for quadratic
constant-width reheating and $3.5$ for quartic reheating through fermion
decay. Kination has a solution with $\Trh>5.96$~MeV only for
$m_\chi>47.2$~GeV. Below this mass the yield at the bound exceeds the
observed abundance by a factor that grows to $4.5$ at $45$~GeV and $11.6$
at $40$~GeV. At $40$~GeV the predicted three-photon flux also exceeds the
INTEGRAL/SPI limit at $N_W=1$. Matching the abundance there would require $N_W<0.09$, outside the $68\%$
interval quoted above. Benchmark B lies close to the line. Benchmark A
has $\delta=0.80$~MeV rather than the $0.92$~MeV of the line at $45$~GeV.
Its coupling is smaller by a factor $2.3$, which accounts for its smaller
excess of $1.95$.

\section{Relation to recent work}
\label{sec:related}
The exothermic recoil interpretation and its target dependence were
developed in Ref.~\cite{deLima:2026shq}. Here two points within the broad
recoil preference share a mediator mass and mixing that are directly
constrained by dark-photon searches. Their dark-matter masses place them
on opposite sides of the $Z$-decay threshold, which changes the
production rate and the histories that reproduce the abundance.
Kim, Kim and Ko~\cite{Kim:2026euj} study an explicit dark Higgs and the
excited fraction for thermal relics motivated by the same event. Here the
abundance is built up during finite reheating in the heavy-radial-scalar limit.

A radiation-dominated calculation starting at $\Trh$ includes only
production after reheating
\cite{deLima:2026shq,Cabo-Almeida:2026uqw,He:2026idw,Choudhury:2026aao}.
We also include production between $\Tmax$ and $\Trh$ and the dilution
of those particles by subsequent entropy release. The latter matters
when comparing with calculations that modify only $H(T)$
\cite{Barman:2026omh}.
Finite reheating and nonstandard expansion have been studied for
Boltzmann-suppressed ultraviolet and electroweak freeze-in
\cite{Bernal:2025fcl,Bernal:2026clv,
Bernal:2026vbg}.
LZ-motivated heavy-Higgsino studies also connect the abundance to
low reheating temperatures and high-scale physics
\cite{Langhoff:2026ujr,Pallis:2026jdc,Hor:2026voa,Unwin:2026qep}.
Finite reheating has also been considered for elastic dark-photon
relics~\cite{Han:2026bmb}.
State-population evolution in standard radiation domination and Galactic
production provide other mechanisms
\cite{Baer:2026fpy,An:2026pkc,Xing:2026civ,Zhu:2026dag}.
Thermal exothermic vector dark matter~\cite{Gemmell:2026yaw},
exothermic scalar dark matter~\cite{Tang:2026lgu}, and transition
charge-radius or anapole freeze-in~\cite{Jodlowski:2026wjx} involve different
production or decay relations.
Other inelastic realizations include scalar dark matter with a dark
photon~\cite{Das:2026buc}, quarkophilic scalar portals
\cite{De:2026win}, and self-interacting pseudo-Dirac dark matter
linked to a Dirac inverse seesaw~\cite{Das:2026uyy}.
Atomic hyperfine excitation provides a different origin for an
endothermic gap and predicts strong target and halo dependence
\cite{Chanda:2026wdl}.
The recoil spectrum also distinguishes boosted dark matter from
inelastic scattering~\cite{Mahapatra:2026glu}.

\section{Conclusions}
\label{sec:conclusions}

We have studied exothermic pseudo-Dirac dark matter with a $0.20$~GeV
dark photon as an interpretation of the $248$~keV LZ event, at two
benchmarks normalized to one expected recoil in the LZ window. The mixing
$\epsilon=8\times10^{-5}$ is consistent with the dark-photon searches considered here,
and both three-photon lifetimes exceed the INTEGRAL/SPI limit.
Self-conversion and reannihilation are negligible. In the static-particle estimate, scattering on the
bath shifts $f_H$ from $1/2$ by less than one percent.

At fixed masses and excited fraction, the recoil rate fixes
$\alpha_D\epsilon^2$, the same coupling product that controls freeze-in
from Standard Model fermions. For $\Tmax=m_\chi$ and thermal dark photons,
quadratic constant-width reheating gives $\Trh=3.0$ and $7.0$~GeV at A
and B, respectively. The corresponding instantaneous-reheating values
are $5.70$ and $11.67$~GeV. For quartic reheating through fermion
decay, we find $91$ and $259$~MeV. The $Z$ pole raises the kination yield at A by a factor of
$18.4$. For $\Tmax=m_\chi$ and $\Trh=5.96$~MeV this leaves an
excess of $1.95$ at the nominal normalization, comparable to the
uncertainty of a one-event rate, while kination at B is viable near
$23$~MeV for thermal dark photons. Along the line $E_0^{\rm Xe}=248$~keV
at one expected recoil, the kination solution crosses the adopted floor near
$m_\chi=47.2$~GeV. Electroweak exchange must therefore be included when an
exothermic recoil signal is used to constrain reheating.

Argon provides a direct test of these benchmarks. Before detector
acceptance, the predicted rates are $1.8$ and $2.8$ events per
tonne-year at A and B, respectively, almost all between $250$ and
$600$~keV. A recoil spectrum on a second target would also help
separate $m_\chi$ from $\delta$ through the dependence of the recoil
energy on the nuclear mass.

The same argument applies to other freeze-in models in which the
interaction measured in nuclear recoils dominates dark-matter
production. Once the masses and the production rate are constrained,
the observed abundance gives one relation among
$(\Trh,\omega,\alpha,\Tmax)$. When most production occurs well above
$\Trh$, Eq.~\eqref{eq:pexp} gives $Y\propto A_{\cal C}\Trh^p$ at
fixed $(\omega,\alpha,\Tmax)$, where $A_{\cal C}$ is the overall
normalization of the production rate. At fixed abundance,
$\Trh\propto A_{\cal C}^{-1/p}$. Thus a factor-of-two change in the
production rate changes the inferred $\Trh$ by only about $5\%$ for
quartic reheating through scalar decay ($p=15$), but by a factor of
two for kination ($p=1$).

\section*{Acknowledgments}
OpenAI's ChatGPT assisted with manuscript editing, literature searches, and implementation and checking of the numerical calculations.

\appendix
\section{Collision-term normalization and numerical checks}
\label{app:source}
For relative momentum $p$, the Born transfer kernel is proportional to
$\sqrt{p^2+b}/[(a+b)^2+4ap^2]$, where $a=m_{\Aprime}^2$ and
$b=2m_\chi\delta$. Its stationary point would occur at
$p^2=(a^2-6ab+b^2)/(4a)$. This is negative at both benchmarks, so the
unsymmetrized kernel is largest at zero incident speed. A factor of two
envelope covers identical-fermion symmetrization relative to the
zero-speed expression used here.

For massless initial particles, quantum statistics can be implemented
without a multidimensional thermal integral. Expanding each distribution
as $f(E)=\sum_{n\geq1}c_ne^{-nE/T}$ gives
\begin{equation}
 {\cal C}_{\rm quantum}(T)=\sum_{n,m\geq1}c_nc_m
 {\cal C}_{\rm MB}\!\left({T\over\sqrt{nm}}\right),\qquad
 c_n=\begin{cases}1&\text{bosons},\\(-1)^{n+1}&\text{fermions}.\end{cases}
 \label{eq:quantum}
\end{equation}
To see the temperature rescaling, transform the two incoming massless momenta by reciprocal factors $\sqrt{n/m}$ and $\sqrt{m/n}$.
Their invariant product and phase-space Jacobian are unchanged, while
the two Boltzmann exponents acquire a common $\sqrt{nm}/T$.
We group the sum by $nm$ and test its convergence numerically.
Final-state dark occupation numbers are negligible.

For the top-quark MB source, the vector and axial weights acquire
$\beta_t(1+2m_t^2/s)$ and $\beta_t^3$, respectively, where
$\beta_t=\sqrt{1-4m_t^2/s}$ and $s\geq4m_t^2$.
The neglected initial-state Fermi correction to this small contribution
is suppressed by $e^{-m_t/T}$.
The $WW$ and $Zh$ polarization sums can be written with virtual-vector
partial widths. In the convention
$\sigma_{b,\mathrm{eff}}=\beta_{\mathrm{in}}^2\sum_{\mathrm{pol}}\sigma_b/4$,
both channels use the source prefactor of Eq.~\eqref{eq:Cff}.
Their effective cross section divided by $\alpha_D\epsilon^2$ is
\begin{equation}
 {\sigma_{b,\mathrm{eff}}\over\alpha_D\epsilon^2}
 ={4\pi\over\sqrt s}|D_b|^2\Gamma_b^{(1)}(\sqrt s)
 (s+2m_\chi^2)\beta_\chi.
\end{equation}
For $WW$,
\begin{equation}
 D_{WW}={m_Z^2\over m_Z^2-m_{\Aprime}^2}(D_A-D_Z),\qquad
 \Gamma_{WW}^{(1)}={\alpha_{\rm em}s^{5/2}\over48m_W^4}
 \beta_W^3(1+20r_W+12r_W^2),\quad r_W={m_W^2\over s}.
\end{equation}
The propagator difference implements the electroweak gauge cancellation.
For $Zh$, let
$\lambda=[1-(m_Z+m_h)^2/s][1-(m_Z-m_h)^2/s]$. Then
\begin{align}
 D_{Zh}&=\tan\theta_W\,{e m_Z\over s_Wc_W}
 {m_Z^2D_Z-m_{\Aprime}^2D_A\over m_Z^2-m_{\Aprime}^2},\\
 \Gamma_{Zh}^{(1)}&={\sqrt\lambda\over48\pi\sqrt s}
 \left[2+{(s+m_Z^2-m_h^2)^2\over4sm_Z^2}\right].
\end{align}
The corresponding BE correction is bounded by replacing each massive
initial distribution with its MB distribution divided by
$1-e^{-m_i/T}$. Its integrated effect is below $0.8\%$ of the total yield
for the histories of Table~\ref{tab:histories}.

\section{Thermodynamic inputs and numerical checks}
\label{app:numerics}

For the Standard Model contributions to $g_s(T)$ and $g_\rho(T)$, we
interpolate the Gondolo--Gelmini table for $T_{\rm QCD}=150$~MeV
\cite{Gondolo:1990dk}, using selected entries from
\texttt{gstar/std.tab} in FreezeIn~\cite{FreezeInStdTable}. When the
dark photon is in thermal equilibrium, we include all three
polarizations in $g_s$ and $g_\rho$.

For the reference point of Ref.~\cite{deLima:2026shq},
$m_\chi=45$~GeV, $\delta=1$~MeV, $m_{\Aprime}=1$~GeV and
$C_{\rm LZ}=1.7\times10^{-17}$, we obtain $1.038$ recoils in the
xenon signal window, in agreement with the quoted $1.04$.

Refining the numerical integration changes the yields at the
temperatures in Table~\ref{tab:histories} by less than $10^{-4}$ in
relative terms. We also check the light-fermion production rate by
integrating separately across the $Z$ resonance. The two evaluations
agree to better than $10^{-4}$ at the temperatures tested.

\bibliographystyle{JHEP}
\bibliography{references}
\end{document}